%% file: main.tex
\documentclass[
reprint,
superscriptaddress,
amsmath,amssymb,
apl,
]{revtex4-2}

\usepackage[utf8]{inputenc}
\usepackage{graphicx,import} 
\graphicspath{{Figures/}}
\usepackage{verbatim}
\usepackage{dcolumn}
\usepackage{bm}
\usepackage[colorlinks]{hyperref}
\usepackage[all]{hypcap}
\usepackage[capitalise,nameinlink]{cleveref} 
\newcommand{\crefadd}[2]{\hyperref[#1]{\cref{#1}#2}}
\newcommand{\Crefadd}[2]{\hyperref[#1]{\Cref{#1}#2}}

\usepackage{tipa} 

\usepackage{xcolor}
\usepackage{braket}
\usepackage{amsmath}
\usepackage{amssymb}
\usepackage{siunitx}
\usepackage{lipsum}
\usepackage{multirow}
\usepackage{transparent}
\usepackage{amssymb,amsmath,amsthm,enumitem}

\definecolor{bluegray}{RGB}{40,180,160}
\definecolor{navygray}{RGB}{110,140,170}
\definecolor{meadowgreen}{RGB}{0,128,0}
\definecolor{coolbrown}{RGB} {165,42,42}

\DeclareSIUnit{\sq}{\Box}

\newcommand{\fr}{f_\mathrm{r}}
\newcommand{\fq}{f_\mathrm{q}}
\newcommand{\Ej}{E_\mathrm{J}}
\newcommand{\futurecitation}[1]{{\color{red}[0]}}

\DeclareUnicodeCharacter{2212}{-}
\hypersetup{
citecolor={bluegray}, 
linkcolor={navygray},
urlcolor={bluegray},
}

\makeatletter
\newcommand*{\balancecolsandclearpage}{%
  \close@column@grid
  \cleardoublepage
  \twocolumngrid
}
\makeatother

\begin{document}
\title{Kinetic inductance coupling for circuit QED with spins}

\author{S.~Günzler}
\email{simon.guenzler@kit.edu}
\affiliation{PHI,~Karlsruhe~Institute~of~Technology,~76131~Karlsruhe,~Germany}
\affiliation{IQMT,~Karlsruhe~Institute~of~Technology,~76131~Karlsruhe,~Germany}

\author{D.~Rieger}
\thanks{First two authors contributed equally.}
\affiliation{PHI,~Karlsruhe~Institute~of~Technology,~76131~Karlsruhe,~Germany} 

\author{M.~Spiecker}
\affiliation{PHI,~Karlsruhe~Institute~of~Technology,~76131~Karlsruhe,~Germany}
\affiliation{IQMT,~Karlsruhe~Institute~of~Technology,~76131~Karlsruhe,~Germany}

\author{T.~Koch}
\affiliation{PHI,~Karlsruhe~Institute~of~Technology,~76131~Karlsruhe,~Germany}

\author{G.~A.~Timco}
\affiliation{Photon~Science~Institute~and~School~of~Chemistry, The~University~of~Manchester, Manchester, UK}
\author{R.~E.~P.~Winpenny}
\affiliation{Photon~Science~Institute~and~School~of~Chemistry, The~University~of~Manchester, Manchester, UK}

\author{I.~M.~Pop}
\affiliation{PHI,~Karlsruhe~Institute~of~Technology,~76131~Karlsruhe,~Germany}
\affiliation{IQMT,~Karlsruhe~Institute~of~Technology,~76131~Karlsruhe,~Germany}
\affiliation{Physics~Institute~1,~Stuttgart~University,~70569~Stuttgart,~Germany}

\author{W.~Wernsdorfer}
\email{wolfgang.wernsdorfer@kit.edu}
\affiliation{PHI,~Karlsruhe~Institute~of~Technology,~76131~Karlsruhe,~Germany}
\affiliation{IQMT,~Karlsruhe~Institute~of~Technology,~76131~Karlsruhe,~Germany} 

\date{\today}

\begin{abstract}
In contrast to the commonly used qubit resonator transverse coupling via the $\sigma_{xy}$-degree of freedom, longitudinal coupling through $\sigma_z$ presents a tantalizing alternative: it does not hybridize the modes, eliminating  Purcell decay, and it enables quantum-non-demolishing qubit readout independent of the qubit$\--$resonator frequency detuning.
Here, we demonstrate longitudinal coupling between  a \{Cr$_7$Ni\} molecular spin qubit ensemble  and the kinetic inductance of a granular aluminum superconducting microwave resonator. 
The inherent frequency-independence of this coupling allows for the utilization of a \qty{7.8}{\giga\hertz} readout resonator to measure the full \{Cr$_7$Ni\} magnetization curve spanning \qtyrange{0}{600}{\milli\tesla}, corresponding to a spin frequency range of $f_\text{spin}=\qtyrange{0}{15}{\giga\hertz}$.
For \qty{2}{\giga\hertz} detuning from the readout resonator, we measure a $1/e$ spin relaxation time $\tau= \qty{0.38}{\second}$, limited by phonon decay to the substrate.
Based on these results, we propose a path towards longitudinal coupling of single spins to a superconducting fluxonium qubit. 
\end{abstract}

\maketitle

Spin systems are a promising platform for quantum information processing (QIP) due to their intrinsic isolation from the environment, which offers long coherence times~\cite{Burkard2023Jun, Zhong2015Jan, Serrano2022Mar, O'Sullivan2024Oct}.
However, this isolation also poses a significant challenge for spin manipulation and readout. 
Superconducting microwave circuits provide a compelling solution within hybrid quantum architectures~\cite{Kurizki2015Mar, Xiang2013Apr, Clerk2020Mar}, combining their versatile design, quantum-non-demolishing (QND) readout, and efficient manipulation with the long coherence times of spin systems. 
Over the past decade, superconducting microwave resonators have demonstrated strong coupling to spin ensembles in various systems, including vacancy centers in diamond~\cite{Kubo2010Sep}, donor atoms in crystalline substrates~\cite{Schuster2010Sep,Bienfait2016Mar, Eichler2017Jan}, and molecular magnets~\cite{Bonizzoni2017Oct, Mergenthaler2017Oct,Gimeno2020Jul, Moreno-Pineda2021Sep}.
Despite these achievements, inhomogeneous broadening limits ensembles coherence, and QIP requires precise control over individual spins - a milestone recently achieved through the readout and manipulation of single electron spins~\cite{Wang2023Jul, O'Sullivan2024Oct}.

These advances rely on transverse coupling between spins and microwave resonators.
For the simplest case of a two-level spin system, the resonator microwave field couples to the spin transverse component: $H_\text{int}^\perp=\hbar g_x \sigma_x (a^\dagger + a)$, where $g_x$ is the coupling constant, $a^\dagger$ and  $a$ are the resonator bosonic operators and $\sigma_x$ is the spin Pauli matrix. 
While mode hybridization due to transverse coupling enables Purcell-effect spin readout~\cite{Bienfait2016Mar, Bienfait2016Purcell, Wang2023Jul, O'Sullivan2024Oct, Eichler2017Jan, Bonizzoni2017Oct, Mergenthaler2017Oct}, it is restricted to the cavity frequency and inherently non-QND~\cite{Blais2021May, Dumas2024Oct}.
Longitudinal interaction provides a robust alternative because the spins couple directly to the resonator frequency via $\sigma_z$, i.e. $H_\text{int}^\parallel = \hbar g_z  \sigma_z a^\dagger a$.
The longitudinal coupling strength $g_z=\delta\omega/2$ leads to a spin-flip induced frequency shift $\delta\omega$.
Since $H_\text{int}^\parallel$ commutes with $\sigma_z$, it mitigates the limitations of transverse coupling, enabling frequency-independent and QND readout~\cite{Didier2015Nov, Richer2016Apr, Dassonneville2020Feb, Chapple2024Dec}.

We argue that kinetic inductance presents a promising platform for longitudinal spin coupling, as it can be modulated by magnetic fields or screening currents induced by nearby spins.
Superconducting quantum interference devices (SQUIDs) are a potential implementation, where the spin magnetic moment tunes the SQUID critical current~\cite{Moreno-Pineda2021Sep, Wernsdorfer2009May} and the corresponding kinetic inductance. 
However, SQUID-based designs face challenges, including limited spin-flux coupling in loop geometries — preventing scalability toward single-spin readout~\cite{Wernsdorfer2009May} — and the intrinsic nonlinearity of Josephson junctions, which limits the power dynamic range of microwave devices.
Disordered superconductors, such as NbN, NbTiN, InOx, and granular aluminum (grAl), provide a more versatile kinetic inductance for longitudinal coupling~\cite{Kroll2019Jun, Samkharadze2016Apr, Bahr2024Mar, Charpentier2025Jan, Borisov2020, Grunhaupt2018Sep, Siddiqi2021Oct, Deutscher1973Jan}. 
These materials offer several key advantages: (i)  intrinsic magnetic field susceptibility without requiring a loop geometry, (ii) resilience in Tesla magnetic fields, and (iii) adaptable nonlinearity, which can be significantly lower compared to JJ-based devices~\cite{Maleeva2018Sep}. 
While kinetic inductance magnetometers (KIMs)~\cite{Luomahaara2014Sep, Asfaw2018Oct, Sypkens2021Feb} represent an important step forward, the longitudinal coupling of spins to a microwave device has yet to be demonstrated.

\input{Figures/Figure1}
In this work, we demonstrate longitudinal coupling between a \{Cr$_7$Ni\} molecular spin qubit ensemble and the kinetic inductance of a grAl microwave resonator.
The spin ensemble polarization is encoded in the resonator frequency, independent of the spin-resonator detuning, enabling us to perform a full microwave readout of the magnetization curve.
Moreover, we demonstrate time resolved measurement of spin ensemble  excitation and decay at \unit{\giga\hertz} detuning from the readout.
Finally, we propose a scheme to achieve single spin longitudinal coupling to a superconducting fluxonium qubit.

\Cref{fig:sample} illustrates the conceptual implementation of longitudinal coupling between a microwave resonator and a spin ensemble. 
The magnetic molecules are grown in a single crystal of chemically identical units, composed of eight chromium atoms with one substituted by nickel, which resemble cyclic wheels (cf. bottom right inset in \crefadd{fig:sample}{a}). 
Each molecule has an effective spin $S=1/2$ and gyromagnetic factor $g\approx1.8$~\cite{Larsen2003Jan, Ardavan2007Jan, Timco2009Mar}, chosen with $g\neq2$ in order to distinguish from spurious coupling to  electronic spins in the environment of superconducting devices~\cite{Borisov2020, Gunzler2025Jan, Samkharadze2016Apr, Kroll2019Jun}.
The molecular crystal is placed at the center of a grAl stripline resonator with a resonance frequency $f_\mathrm{r}= \qty{7,8}{\giga\hertz}$ and internal quality factor $Q_\mathrm{i}>10^5$ (cf.~\cref{sec:supp:grAlInField}). 
The grAl film has a kinetic inductance fraction $\alpha\approx 1$~\cite{Grunhaupt2018Sep} (sheet inductance $L_\square = \qty{0,9}{\nano\henry\per\sq}$) which arises from its microstructure, effectively forming a 3D network of JJs~\cite{Deutscher1973Jan, Gralmonium, Maleeva2018Sep}.
To drive the spin ensemble using microwave transverse fields, on the same chip (cf.~\crefadd{fig:sample}{b,c}), we pattern two low impedance resonators  ($f_{\text{Nb},1}=\qty{9,8}{\giga\hertz}$, $f_{\text{Nb},2}=\qty{11,2}{\giga\hertz}$) made of niobium, which can sustain relatively high current densities. 
\input{Figures/Figure2}

The longitudinal spin-resonator interaction arises from the coupling of the spin magnetic moments ${\mu=-g\mu_\mathrm{B} \sigma_z/2 }$ to the resonator's kinetic inductance $L_\text{kin}$, where $\mu_\mathrm{B}$ denotes the Bohr magneton.
The spin ensemble magnetization $M$ increases $L_\text{kin}$ and shifts the resonator frequency $\fr$ by $\delta f_\mathrm{M}$:
\begin{align}
    \delta f_\mathrm{M}/f_{\mathrm{r},0}  \propto -M^2\,. \label{eq:dfM_shift}
\end{align}
This frequency shift results from persistent currents $I_\text{p}$ induced by the perpendicular component of the crystal field $B_{\uparrow,\perp}\propto M$.
These currents locally increase the kinetic inductance $L_\mathrm{kin}(I_\mathrm{p})\approx L_{\mathrm{kin},0} (1+ (I_\mathrm{p}/I^*)^2)$, where $I^*$ characterizes the grAl nonlinearity and is typically on the order of the critical current~\cite{Luomahaara2014Sep, HoEom2012Aug}.
Given that the grAl effective penetration depth exceeds the resonator width by orders of magnitude~\cite{Abraham1978Sep, Gershenson1982Apr}, we consider $I_\text{p}\propto B_{\uparrow,\perp}$.
Note that the contribution of the superconducting gap suppression to $\delta f_\mathrm{M}$ is neglected, as the persistent current suppression dominates by 3 orders of magnitude for our resonator geometry~\cite{Borisov2020}.

Using the grAl resonator, in \cref{fig:magnetization}, we demonstrate the measurement of the \{Cr$_7$Ni\} crystal magnetization curve  in parallel magnetic field $B_\parallel$.
The observed frequency shift $\delta\fr(B_\parallel,M)$ (cf.~\crefadd{fig:magnetization}{a}) consists of two contributions:
the shift $\delta f_\mathrm{M}$ from the crystal magnetization $M$ (cf.~\cref{eq:dfM_shift}) and an additional parabolic frequency decrease due to the grAl gap suppression in $B_\parallel$~\cite{Borisov2020}.
To disentangle these effects, we fit the parabolic tail of the frequency shift $\delta\fr(B_\parallel, M_\mathrm{S})$  in the field range where the ensemble magnetization is saturated ($M=M_\text{S}$), specifically for $B_\parallel > \SI{0.32}{\tesla}$ in \crefadd{fig:magnetization}{a}.
Subtracting the fitted parabolic response $\delta\fr(B_\parallel, 0)$ (dashed line in \crefadd{fig:magnetization}{a}) from the measured data $\delta\fr(B_\parallel, M)$ isolates the contribution $\delta f_\mathrm{M}$ from the crystal’s magnetization, shown in \crefadd{fig:magnetization}{b}. 
In addition, due to the resonator's geometric inductance, near $B_\parallel = \qty{0,29}{\tesla}$ we observe small transverse couplings  to the $g=1.83$ spins and to spurious $g=2$ spin impurities.

The magnetization curve in \crefadd{fig:magnetization}{b}, extracted from $\delta f_\mathrm{M}$ using \cref{eq:dfM_shift},  increases from $M=0$ to saturation $M=M_\mathrm{S}$ in fields above $\qty{300}{\milli\tesla}$.
It follows the paramagnetic response $M=M_\mathrm{S}\tanh(\frac{g\mu_\mathrm{B} B_\parallel}{2k_\mathrm{B}T_\mathrm{S}})$ for a spin $1/2$ ensemble, from which we fit a spin temperature of $T_\mathrm{S}=\qty{70}{\milli\kelvin}$ ($k_\mathrm{B}$ is the Boltzmann constant).
Notably, $T_\mathrm{S}$ is above the cryostat temperature of $T\approx\qty{30}{\milli\kelvin}$, indicating limited thermalization of the molecular spin crystal via the vacuum grease (cf.~\cref{fig:sample}).
In a subsequent run with a smaller crystal, we extract a spin temperature of $T_\mathrm{S}=\qty{30}{\milli\kelvin}$ (cf.~\cref{sec:supp:Nb_magnetization_curve}), illustrating the challenge of thermalizing the ensemble - an issue commonly referred to as the phonon bottleneck~\cite{Chiorescu2000Apr}.

\input{Figures/Figure3}
To conclude the spin ensemble characterization, we leverage the detuning-independent nature of kinetic inductance readout, which also remains effective during spin manipulation.
As shown in \crefadd{fig:excitation}{a} , we perform two-tone spectroscopy of the spins around \qty{9,85}{\giga\hertz}, $\qty{2}{\giga\hertz}$ detuned from the readout resonator, and we measure $g=1.83\pm0.01$ (black dashed line), comparable to values reported in the literature~\cite{Larsen2003Jan, Timco2009Mar}.
The excitation pattern reflects the expected convolution of the frequency dependent drive amplitude, shaped by the Nb resonators Lorentzian response, combined with the spin ensemble’s inhomogeneous linewidth.

On resonance with the Nb drive (red marker in \crefadd{fig:excitation}{a}), we measure time-resolved excitation and decay of the ensemble magnetization (cf.~\crefadd{fig:excitation}{b,c}).
We observe a non-exponential decay with $1/e$ reached after $\tau= \qty{0.38}{\second}$.
This lifetime, significantly longer than reported for the same \{Cr$_7$Ni\} molecules in solution~\cite{Ardavan2007Jan}, is likely limited by the decay of the crystal phonons into the substrate (phonon bottleneck~\cite{Chiorescu2000Apr}), consistent with Ref.~\cite{Wernsdorfer2005Aug}.
Note that we exclude drive induced heating as origin of the spin excitation, since the ensemble magnetization remains unchanged when the spins are detuned from the drive (e.g. at $B_\parallel=\qty{0,36}{\tesla}$ in \crefadd{fig:excitation}{a}).
\input{Figures/Figure4}

In \cref{fig:spin_coupling}, we propose a kinetic inductance coupling scheme for a single spin, utilizing established grAl circuit elements.
To obtain longitudinal coupling to the highly localized dipole field of a single spin, we need to concentrate the kinetic inductance within a nanoscopic volume.
This volume reduction inherently increases device nonlinearity~\cite{Maleeva2018Sep}, which for the \qty{20}{\nano\meter} cross-section in \crefadd{fig:spin_coupling}{a} effectively results in a Josephson junction, referred to as a grAl nanojunction~\cite{Gralmonium}.
The spin dependent local magnetic field threading the nanojunction locally induces a quadratic suppression of the grAl superconducting gap~$\Delta$, resulting in a spin-state-dependent Josephson energy $\Ej:\,\Ej^{\uparrow} \neq \Ej^{\downarrow}$.

We propose embedding the grAl nanojunction into a fluxonium quantum circuit operated at the half-flux sweet spot, where the qubit frequency is exponentially sensitive to variations in $\Ej$ ~\cite{Gralmonium}. 
The resulting circuit shown in \crefadd{fig:spin_coupling}{b} can be described by the standard fluxonium Hamiltonian 
\begin{align}
    H_\mathrm{flux} &= 4 E_\text{C} n^2 + \frac{1}{2}E_\text{L}\left(\varphi - 2\pi\frac{\Phi_\text{ext}}{\Phi_0}\right)^2 - E_\text{J}\cos\varphi\,,
    \label{eq:fluxoniumHamiltonian}
\end{align}
where $E_\text{L}=(\Phi_0 / 2\pi)^2 / L_\text{q}$ and $E_\text{C}=e^2 / 2 C$ denote the fluxonium inductive and charging energies, respectively.
Besides the trivial modulation from the bias field $B_\parallel$, $\Ej=\Ej(B_\parallel, \sigma_z)$ also depends on the spin qubit state $\sigma_z$, thereby implementing a longitudinal coupling where $\sigma_z$ commutes with $H_\mathrm{flux}$.
A practical realization of such a circuit should be resilient to the high magnetic fields required for spin qubit operation, as recently demonstrated in a gradiometric geometry in Ref.~\cite{Gunzler2025Jan}.

To estimate the fluxonium qubit frequency shift $\delta \fq$ due to a spin flip in a realistic scenario, we model the spin as a point-like magnetic dipole with a magnetic moment $\mu=10 \mu_\mathrm{B}$ oriented (anti-) parallel to the bias field $B_\parallel$.
Such spins can be engineered in single molecule magnets with strong anisotropy, forming an effective two-level-system with large magnetic moment~\cite{Moreno-Pineda2021Sep}.
For a spin centered at a distance $d$ above the nanojunction, we numerically calculate the local grAl gap suppression $\Delta(\Vec{B}_\uparrow(\Vec{r})+\Vec{B}_\parallel)/\Delta(0) = \sqrt{1-(|\Vec{B}_\uparrow( \Vec{r})+\Vec{B}_\parallel|/B_\mathrm{c})^2}$~\cite{Douglass1961Apr} at position $\Vec{r}$ within its $(\qty{20}{\nano\meter})^3$ volume.
By numerically solving \cref{eq:fluxoniumHamiltonian} for the corresponding $\Ej\propto \iiint \mathrm{d}\Vec{r} \Delta(\Vec{r})$, we find the qubit frequency shift $\delta \fq$, shown in \crefadd{fig:spin_coupling}{c}.
Positioning the spin as close as possible to the nanojunction and operating in the $B_\parallel\sim10^2\,\unit{\milli\tesla}$ range enables spin-state-dependent frequency shifts $\delta \fq$ on the order of \unit{\kilo\hertz}.
Given that superconducting qubit frequencies can be resolved with \unit{\kilo\hertz} accuracy~\cite{Burnett2019Jun, Somoroff2023Jun}, we are optimistic that longitudinal coupling via kinetic inductance will enable single shot QND readout of individual spins.

In conclusion, we have demonstrated longitudinal coupling between a \{Cr$_7$Ni\} molecular spin ensemble and a grAl superconducting microwave resonator via kinetic inductance, enabling frequency-independent and non-demolishing spin readout. 
The independence of the readout contrast on detuning between the spins and the resonator allowed us to measure the full magnetization curve over a field range of hundreds of \unit{\milli\tesla} and observe time-resolved spin dynamics at GHz detuning. 
To overcome ensemble limitations like inhomogeneous broadening and the phonon relaxation bottleneck, we propose extending this longitudinal coupling scheme to a single spin positioned in the close vicinity of a grAl nanojunction in a fluxonium qubit.
For state-of-the-art grAl devices, our simulations predict a spin-dependent qubit frequency shift up to \unit{\kilo\hertz}, opening the path for QND interaction at the single-spin level.
Order of magnitude stronger interaction could be possible by miniaturizing the nanojunction by a factor of two and embedding the magnetic molecule in its center.

\section*{Acknowledgements}
We are grateful to Eufemio Moreno-Pineda and Kiril Borisov for fruitful discussions and we acknowledge technical support from S. Diewald and L. Radtke. 
Funding was provided by the German Research Foundation (DFG) via the Gottfried Wilhelm Leibniz-Award (ZVN-2020\_WE 4458-5) and by the European Research Council via project number 101118911 (DarkQuantum). 
Facilities use was supported KIT Nanostructure Service Laboratory (NSL). 
We acknowledge qKit for providing a convenient measurement software framework.

\bibliography{references}
\balancecolsandclearpage

\onecolumngrid
\section*{Appendices}
\vspace{0.6cm}
\twocolumngrid
\appendix
\section{Fabrication details}
The resonators discussed in this manuscript are fabricated on a double-side-polished c-plane sapphire substrate. 
They are patterned using lift-off optical lithography with AZ5214E photoresist, followed by development in AZ developer. 
Before metal deposition, an Ar/O$_2$ plasma cleaning step is performed using a Kaufman ion source. 
Additionally, a titanium gettering step reduces the chamber pressure to $p\sim 10^{-7}\,\unit{\milli\bar}$ before evaporation in a Plassys MEB 550S shadow evaporation system. 
The \qty{20}{\nano\meter} granular aluminum (grAl) film is deposited at \qty{1}{\nano\meter/\second}, yielding a sheet resistance of \qty{1,1}{\kilo\ohm/\sq}. 
In a second lithography step, \qty{50}{\nano\meter} thick niobium drive resonators are patterned and deposited with an evaporation rate of \qty{1}{\nano\meter/\second}, using the same optical procedure.

\section{grAl resonator in magnetic field}
\label{sec:supp:grAlInField}
\input{Figures/FigureSupp_Qi}

In \cref{fig:supp:Qi}, we confirm the resilience of the grAl readout resonator in parallel magnetic field $B_\parallel$.
The internal quality factor remains above $Q_\mathrm{i}\gtrsim\,10^5$, even when accounting for systematic Fano uncertainty in the loss measurement~\cite{Rieger2023Jul}, except for two distinct dips.
These increased losses occur where the resonator frequency matches the ESR of the $g=1.83$ molecular spin ensemble and of spurious $g=2$ spins, commonly observed in the environment of superconducting devices~\cite{Borisov2020, Gunzler2025Jan, Samkharadze2016Apr, Kroll2019Jun}.
The latter can serve as an in-situ calibration of the on-chip magnetic field strength.

Notably, the quality factors of the grAl resonator are not affected by the \textit{Apiezon~N}~\cite{ApiezonN} vacuum grease on top, used to thermalize and attach the crystal (cf.~\cref{fig:sample}).
Within the uncertainty, the values reported in \cref{fig:supp:Qi} are equivalent to values reported in the literature~\cite{Grunhaupt2018Sep, Borisov2020, Rieger2023Jul} and could be further improved by an order of magnitude for optimized geometry, substrate and cleaning procedures, as shown in Ref.~\cite{Gupta2024Nov}.

Beyond field strength, we also calibrate the field direction to ensure that all measurements in this manuscript are performed with $B_\parallel$ strictly aligned with the substrate plane. 
This prevents hysteretic, non-monotonic resonance frequency shifts and additional perpendicular field ($B_\perp$) effects that would complicate the extraction of $\delta f_\mathrm{M}$.
To compensate for minor chip misalignments within our cylindrical sample holder (identical to Refs.~\cite{Borisov2020, Gunzler2025Jan, Rieger2023Jul}) we introduce an out-of-plane compensation field $B_{\perp,\mathrm{comp}}$ for for each applied $B_\parallel$.
We determine $B_{\perp,\mathrm{comp}}$ by sweeping the perpendicular field component $B_\perp$ and maximizing the resonator frequency.

\section{Additional magnetization curve}
\label{sec:supp:Nb_magnetization_curve}
\input{Figures/FigureSupp_magnetization}
In \cref{fig:supp:magnetization}, we confirm that the presence of the Nb drive resonators, which introduces magnetic field distortions, still allows reliable extraction of the spin ensemble magnetization.
As shown in \cref{fig:sample}, these Nb resonators are patterned directly adjacent to the grAl readout resonator.
The extracted shift $\delta f_\mathrm{M}$ and corresponding magnetization $M$, shown in \crefadd{fig:supp:magnetization}{b}, exhibit saturation behavior similar to the sample without Nb resonators (cf.~\cref{fig:magnetization}). 
The frequency shift at saturation $\delta f_\mathrm{M}(M_\mathrm{S})$ is smaller compared to \cref{fig:magnetization} due to the use of a smaller molecular crystal.
From a fit to the paramagnetic response $M=M_\mathrm{S}\tanh(\frac{g\mu_\mathrm{B} B_\parallel}{2k_\mathrm{B}T_\mathrm{S}})$ for a spin $1/2$ ensemble, we extract a spin temperature of $T_\mathrm{S}=\qty{30}{\milli\kelvin}$.
We attribute the lower spin temperature compared to \cref{fig:magnetization} to the higher surface-to-volume-ratio of the smaller crystal, which enhances phonon decay into the substrate.
For bias fields above $B_\parallel>\qty{300}{\milli\tesla}$, the magnetization remains constant, but an additional shift in $\delta f_\mathrm{M}$ arises due to out-of-plane magnetic fields generated by screening currents in the niobium structures of the drive resonators.

\end{document}

%% file: Figures/Figure1.tex
\begin{figure*}[t!]
\centering
\includegraphics[]{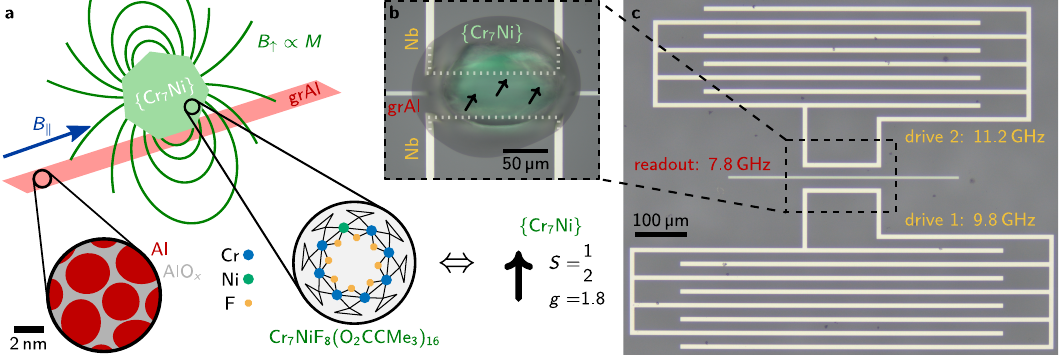}
\caption{
\textbf{Implementation of kinetic inductance coupling between a superconducting resonator and molecular spins.}
\textbf{a}~Sketch of the resonator-spin system: high kinetic inductance granular aluminum (grAl) resonator (red strip) coupled to a \{Cr$_7$Ni\} micro-crystal (green).
Left inset: grAl consists of crystalline aluminum grains (red) in an amorphous AlO$_x$ matrix (gray)~\cite{Deutscher1973Jan}. 
Right inset: the single magnetic molecule $\mathrm{Cr_7NiF_8(O_2CCMe_3)_{16}} \equiv $\{Cr$_7$Ni\}  has effective spin $S=1/2$ and $g=1.8$~\cite{Larsen2003Jan, Ardavan2007Jan}.
The magnetization of the spin ensemble due to in-plane field $B_\parallel$ (blue arrow) creates a magnetic field $B_\uparrow$ (green lines). This field locally enhances the kinetic inductance of the grAl resonator and provides the mechanism for longitudinal coupling.
\textbf{b}~Optical microscope image of the magnetic crystal (green) placed on top of the grAl readout resonator.
The crystal is attached with \textit{Apiezon~N}~\cite{ApiezonN} vacuum grease (transparent gray), which provides thermal conductivity at cryogenic temperatures, low levels of magnetic susceptibility and low microwave losses (cf.~\cref{sec:supp:grAlInField}).
To excite the spins (black arrows) composing the crystal we use inductively coupled niobium lines, visible above and below the grAl resonator and shown as dashed white outlines in the region underneath the crystal.
The \qty{150}{\micro\meter}-long inductor sections of the drive resonators parallel to the central grAl strip generate radio-frequency magnetic fields perpendicular to the spin quantization axis (defined by $B_\parallel$).
\textbf{c}~Zoom-out: The grAl readout resonator (center) is flanked by two low-impedance niobium resonators (top, bottom), necessary to excite the spins. Note that the frequencies $f_{\text{Nb},1}$, $f_{\text{Nb},2}$ of the niobium drive resonators are several \unit{\giga\hertz} detuned from the readout $f_\text{r}$.
}
\label{fig:sample}
\end{figure*}

%% file: Figures/Figure2.tex
\begin{figure*}[t!]
\centering
\includegraphics[]{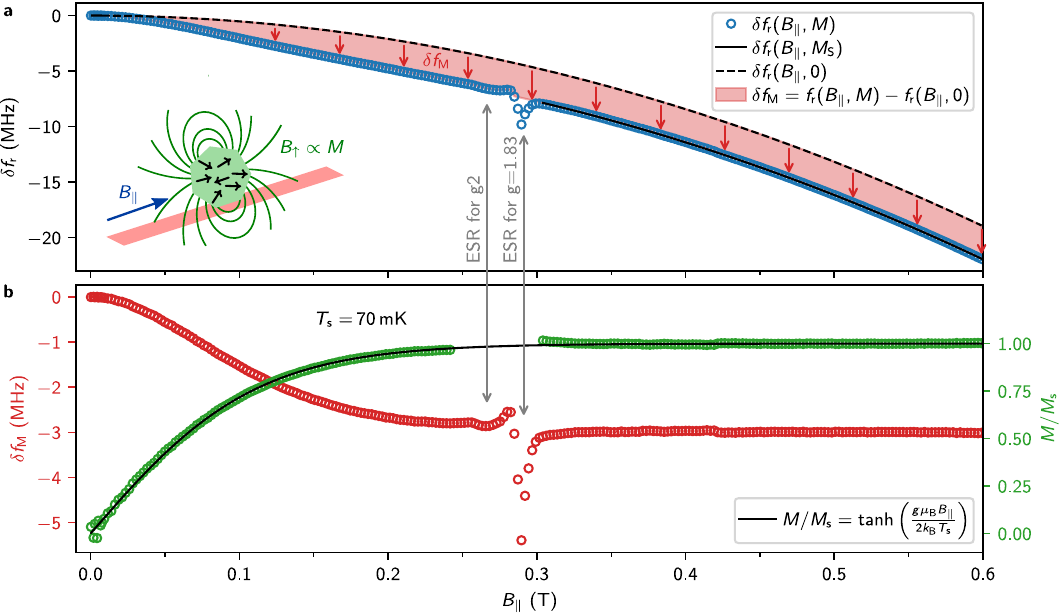}
\caption{\textbf{Detuning independent readout of the spin ensemble magnetization.}
\textbf{a}~Resonance frequency shift of the grAl readout resonator coupled to the molecular spin ensemble $\delta f_\mathrm{r}(B_\parallel,M)$ (blue markers) in magnetic field.
The parabolic frequency shift of the bare resonator $\delta f_\mathrm{r}(B_\parallel,M=0)$ (black dashed line) is given by the suppression of the grAl superconducting gap $\delta(B_\parallel)$ in magnetic field~\cite{Borisov2020}. 
The resonator experiences an additional frequency shift $\delta f_\mathrm{M}$ due to the magnetization of the spin ensemble (red shift), yielding the measured total frequency shift $\delta f_\mathrm{r}(B_\parallel,M)$.
To disentangle these two contributions, we fit a parabolic frequency shift $\delta f_\mathrm{r}(B_\parallel,M_\mathrm{S})$ to the tail of the data for $B_\parallel > \SI{0.32}{\tesla}$ (black solid) where the magnetization of the crystal is saturated $(M=M_\mathrm{S})$. 
We can then remove the vertical offset to get $\delta f_\mathrm{r}(B_\parallel,0)$ (dashed black) and $\delta f_\mathrm{M}$ (red arrows).
At $B_\parallel=\qty{0.29}{\tesla}$, as expected for $g=1.83$, we measure an avoided level crossing due to the transverse coupling of the spin ensemble to the geometric inductance of the resonator.
At $B_\parallel=\qty{0.26}{\tesla}$ we observe an additional feature corresponding to transverse coupling of $g=2$ spurious ESR~\cite{Borisov2020, Gunzler2025Jan, Samkharadze2016Apr, Kroll2019Jun}.
\textbf{b}~Extracted frequency shift $\delta f_\mathrm{M}$ (red markers) due to the  magnetic field of the crystal and corresponding magnetization $M\propto \sqrt{\delta f_\mathrm{M}}$ (green markers).
Above $B_\parallel = \qty{0.32}{\tesla}$, the crystal magnetization saturates, and $\delta f_\mathrm{M}$ remains constant. In the range $B_\parallel = \qtyrange{0.244}{0.302}{\tesla}$, anti-crossings (see double arrow markers) arising from residual transverse coupling to $g=2$ and $g=1.83$ spins prevent a reliable extraction of the magnetization.
Note that the magnetization curve shown here is measured on a sample without Nb drive resonators (cf.~\cref{fig:sample}) in order to avoid field distortions (cf.~\cref{sec:supp:Nb_magnetization_curve}).
}
\label{fig:magnetization}
\end{figure*}

%% file: Figures/Figure3.tex
\begin{figure}[t]
\centering
\includegraphics[scale=1.0]{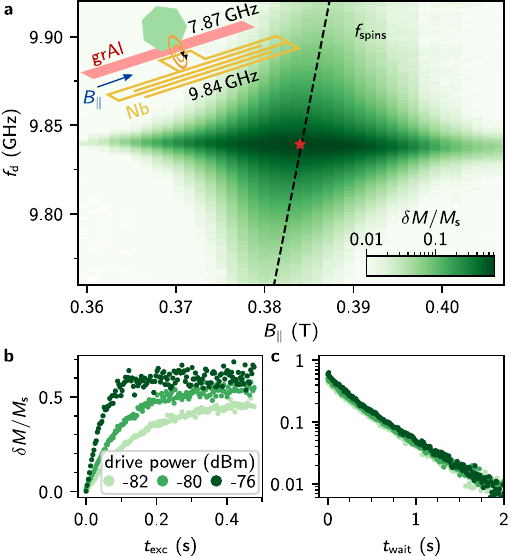}
\caption{\textbf{Excitation and decay of the spin ensemble $\mathbf{\qty{2}{\giga\hertz}}$ detuned from the readout.} 
\textbf{a} Continuous wave two-tone spectroscopy of the spin ensemble: To excite the spins, we sweep a drive tone $f_\mathrm{d}$ in the vicinity of the niobium resonator frequency $f_\text{Nb1} = \qty{9,84}{\giga\hertz}$ while monitoring the readout resonator at $f_\text{r}= \qty{7,87}{\giga\hertz}$. 
The relative change of the ensemble magnetization $\delta M/M_\text{S}$ (green colorbar) is calibrated to the measured shift in  $f_\text{r}$ utilizing a magnetization curve (cf. \crefadd{fig:magnetization}{b}).
The excitation of the spin ensemble is only effective within the bandwith $\kappa = \qty{2}{\mega\hertz} $ of the Nb drive resonator.
The dashed line indicates the center frequency of the spin distribution given by $\text{h}f = g\mu_\text{B} B_\parallel$ with $g=1.83\pm0.01$.
\textbf{b,\,c} Time domain characterization of the spin ensemble at $B_\parallel = \qty{0.384}{\tesla}$ (red marker in \textbf{a}): excitation to saturation (\textbf{b}) and decay from saturation (\textbf{c}) for different drive powers, with a $1/e$ time $\tau= \qty{0.38}{\second}$.
}
\label{fig:excitation}
\end{figure}

%% file: Figures/Figure4.tex
\begin{figure}[t!]
\centering
\includegraphics{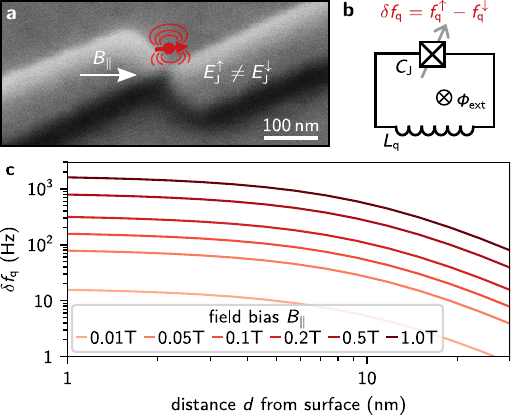}
\caption{\textbf{Towards longitudinal coupling between superconducting circuits and single spins.}
\textbf{a}~Implementation of a spin-modulated Josephson junction: scanning electron microscope image of a grAl nanojunction with a single spin sketched on top. 
The nanojunction consists of a $\approx(\qty{20}{\nano\metre})^3$ grAl volume~\cite{Gralmonium},  which remains coherent in Tesla-scale magnetic fields~\cite{Gunzler2025Jan}.
When biased in magnetic field $B_\parallel$, the two spin orientations generate different local magnetic fields, leading to distinct Josephson energies $\Ej^{\uparrow}$ and $\Ej^{\downarrow}$. 
\textbf{b}~We use the spin-state-dependent nanojunction to build a fluxonium quantum circuit with junction capacitance $C_\mathrm{J}$, superinductor $L_\mathrm{q}$ and external flux $\Phi_\mathrm{ext}$.
The spin polarization tunes the Josephson energy $E_\mathrm{J}$ and, consequently, the fluxonium transition frequency $f_\mathrm{q}$.
\textbf{c}~Simulation of the qubit frequency shift $\delta f_\mathrm{q}$ for a spin-flip versus distance $d$ from the nanojunction surface for several $B_\parallel$ bias values. 
Note that for the calculation we used a spin with magnetic moment $\mu = 10\mu_\mathrm{B}$ (see Ref.~\cite{Moreno-Pineda2021Sep}) on top of a regular gralmonium qubit~\cite{Gralmonium}.
}
\label{fig:spin_coupling}
\end{figure}

%% file: Figures/FigureSupp_Qi.tex
\begin{figure}[t]
\centering
\includegraphics[scale=1.0]{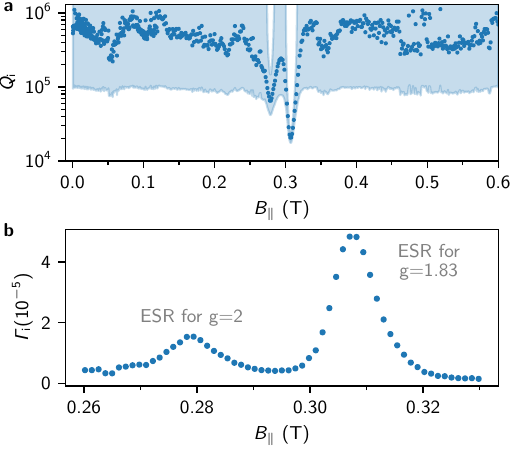}
\caption{
\textbf{Resilience of grAl resonators in parallel field.} 
Internal quality factor $Q_\mathrm{i}$ (blue markers) corresponding Fano uncertainty range~\cite{Rieger2023Jul} (blue shaded area) of the grAl resonator in magnetic field $B_\parallel$.
The resonator is measured at an average photon number $\Bar{n}\approx10$ with a coupling quality factor $Q_\mathrm{c}=\qty{25e3}{}$.
The dips in $Q_\mathrm{i}$, shown as peaks in the decay rate $\Gamma_\mathrm{i}=1/Q_\mathrm{i}$ in (b), correspond to ESR of spurious electronic spins with $g=2$ (left) and the molecular spin ensemble with $g=1.83$ (right).
The associated resonance frequency shifts are presented in \cref{fig:supp:magnetization}.
}
\label{fig:supp:Qi}
\end{figure}

%% file: Figures/FigureSupp_magnetization.tex
\begin{figure*}[bt]
\centering
\includegraphics[]{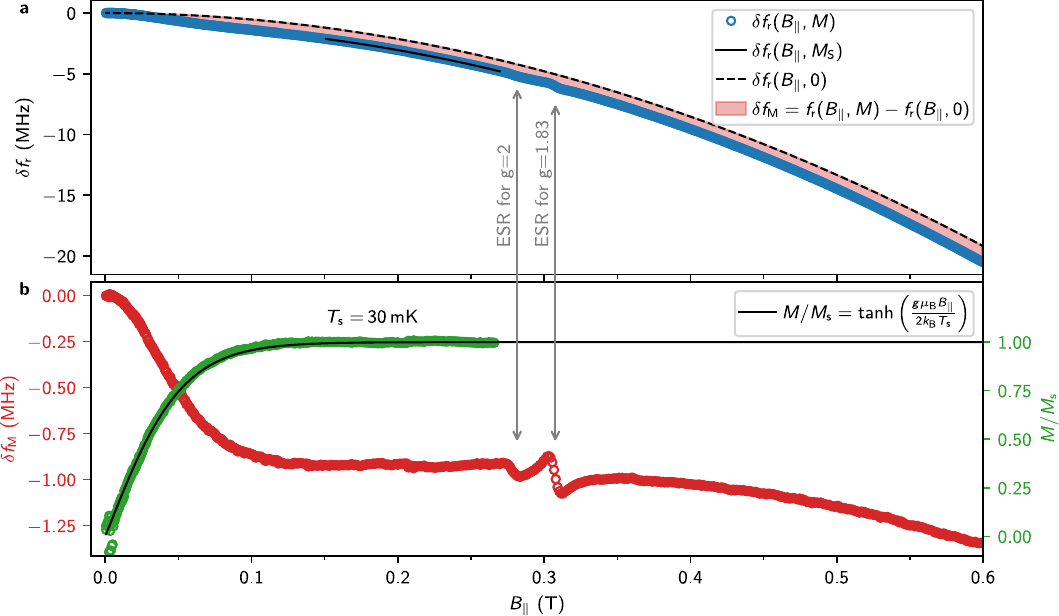}
\caption{
\textbf{Readout of the spin ensemble magnetization in the presence of on‐chip Nb drive resonators.}
\textbf{a}~Resonance frequency shift $\delta f_\mathrm{r}(B_\parallel,M)$ (blue markers) of the grAl readout resonator coupled to the molecular spin ensemble, measured in the presence of Nb drive resonators (which remain undriven during this measurement). 
To extract the magnetization $M\propto \sqrt{\delta f_\mathrm{M}}$ (green markers) in \textbf{b}, we fit the parabolic frequency shift $\delta f_\mathrm{r}(B_\parallel,M_\mathrm{S})$ for saturated spin crystal magnetization $(M=M_\mathrm{S})$ in the range $B_\parallel=\qtyrange{150}{270}{\milli\tesla}$ and follow the main text procedure (cf.~\cref{fig:magnetization}).
For fields above $B_\parallel = \qty{0.27}{\tesla}$, as shown by the frequency shift $\delta f_\mathrm{M}$ (red markers) in \textbf{b}, anti-crossings arising from residual transverse coupling to $g=2$ and $g=1.83$ spins (gray arrows) as well as screening fields from the superconducting Nb structures distort the extracted frequency shift $\delta f_\mathrm{M}$.
}
\label{fig:supp:magnetization}
\end{figure*}